\documentclass[pra, twocolumn, superscriptaddress]{revtex4-2}
\usepackage{xcolor}
\usepackage{todonotes}
\usepackage{amsmath,amssymb,bm}
\usepackage{physics}
\usepackage{upgreek}
\usepackage{graphicx}
\usepackage{dcolumn}
\usepackage{dsfont}
\usepackage{nccmath}
\usepackage[colorlinks=true, allcolors=blue]{hyperref}
\usepackage{amsthm}
\usepackage{units}
\usepackage{mathrsfs}
\usepackage{hhline}
\usepackage{subfigure}
\usepackage{svg}
\usepackage{soul}
\usepackage{enumitem}
\setlist[enumerate]{itemsep=0mm}
\usepackage[normalem]{ulem}
\begin{document}
\title{Robust Macroscopic Schr\"odinger's Cat on a Nucleus}
\author{Pragati Gupta}
\affiliation{Institute for Quantum Science and Technology, University of Calgary, Alberta T2N 1N4, Canada}
\author{Arjen Vaartjes}
\affiliation{School of Electrical Engineering and Telecommunications, UNSW Sydney, Sydney, New South Wales 2052, Australia}
\author{Xi Yu}
\affiliation{School of Electrical Engineering and Telecommunications, UNSW Sydney, Sydney, New South Wales 2052, Australia}
\author{Andrea Morello}
\affiliation{School of Electrical Engineering and Telecommunications, UNSW Sydney, Sydney, New South Wales 2052, Australia}
\author{Barry C. Sanders}
\affiliation{Institute for Quantum Science and Technology, University of Calgary, Alberta T2N 1N4, Canada}
\begin{abstract}
    We propose a scheme to generate spin cat states,
    i.e., superpositions of maximally separated quasiclassical states on a single high-dimensional nuclear spin in a solid-state device. 
    We exploit a strong quadrupolar nonlinearity to drive the nucleus significantly faster than usual gate sequences, achieving 
    collapses and revivals two orders of magnitude faster than the dephasing timescale. 
    Furthermore, these states are engineered without entanglement with an ancilla, hence, are robust against error propagation. 
    With our multitone control, we can realize arbitrary high-spin rotations within an experimentally feasible regime, as well as transform a spin coherent state to a spin cat state using only phase modulation, opening the possibility of storing and manipulating high-fidelity cat states.
\end{abstract}
\maketitle
\section{Introduction}
A Schr\"odinger cat state~\cite{Schrodinger35Nature,YS86PRL,S89PRA} is an equal superposition of two distinct ``classical" states of a system~
\cite{G63PR,p86springer,P72CommunMathPhys},
that
is a probe for fundamental aspects of quantum physics, including the measurement problem~\cite{S05RevModPhys}, macro-realism~\cite{L80ProgTP,L02JoP}, quantum-classical transition~\cite{BLS+13RevModPhys}, and non-locality~\cite{W00NJP}.
Cat states also serve as important resources for quantum error correction~\cite{OPH+16Nature,PGC19PRX,PJG+20ScienceAdv, G21PRL}, quantum sensing~\cite{SJKA10PRA,FDG+16Nature,GLM06PRL}, 
and quantum communication~\cite{KBB02PRA,BRP+10PRL}. 
Thus, practical realization of cat states is of interest for  both foundational and technological applications.

In continuous-variable systems, cat states are generated using nonlinearity induced by light-matter interactions in the Jaynes-Cummings model~\cite{FGB+08Nature,LMJ+17PRA}, e.g., a harmonic oscillator coupled to a superconducting qubit~\cite{GFP20Nature,HRO+17NatCommun,RRM+18Science} or an atom in an optical cavity~\cite{HWD19NatPhoton}.
\textcolor{black}{However, errors occurring in the qubit can propagate to the cavity, corrupting such nonclassical states~\cite{RRM+18Science}.}
Alternatively, Greenberger-Horne-Zeilinger (GHZ) states, which are spin cat states
(SCSs)
for permutationally symmetric systems, can be realized by entangling multiple qubits~\cite{OLK+19Science,RNO+00Science,SXL+17PRL,SXL+19Science}, but such SCSs are fragile---with their decoherence rate proportional to the square of the number of qubits~\cite{MSB+11PRL}. 

\begin{figure}
    \centering
    \includegraphics[width = \linewidth]{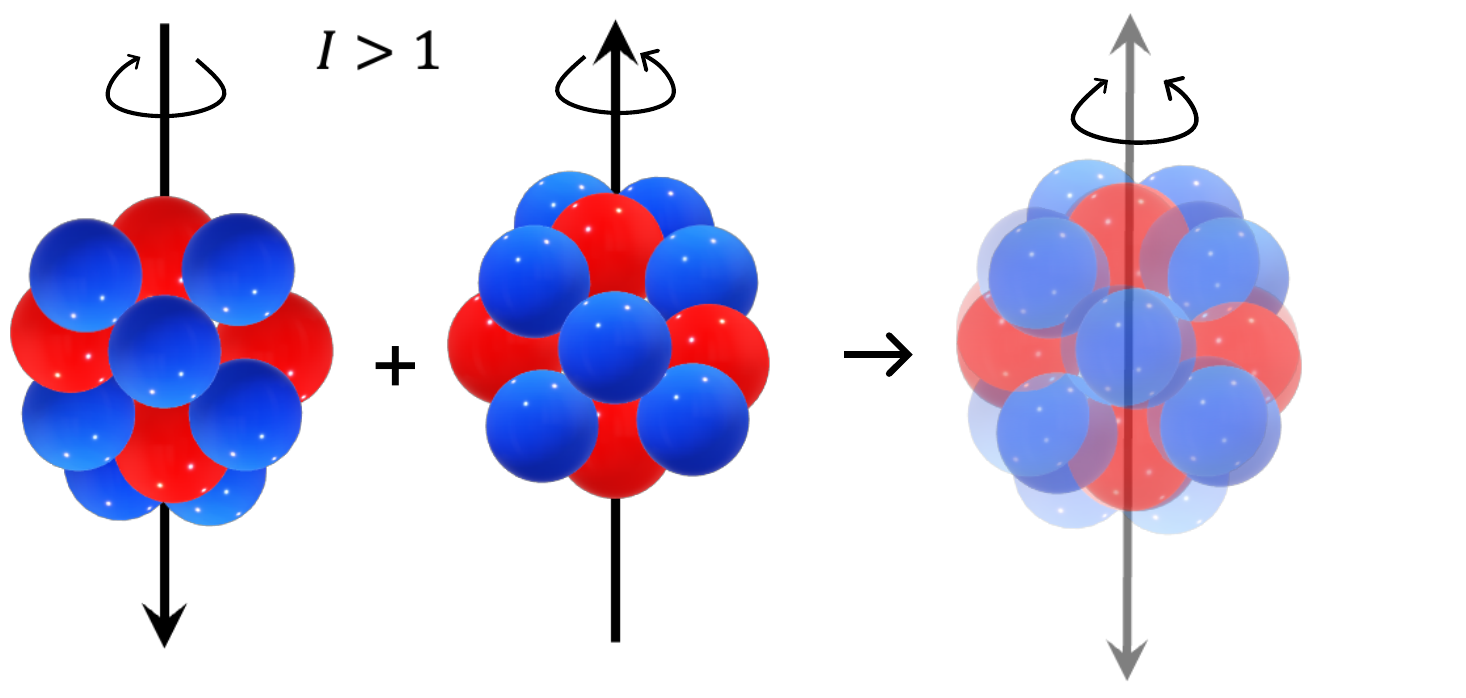}
    \caption{%
    Superposition of spin-up and spin-down instances of a high-spin nucleus \textcolor{black} {($I>1$)},
    comprising protons (red) and neutrons (blue).}
    \label{fig:nuclearcat}
\end{figure}

\textcolor{black}{We propose a new platform to generate cat states deterministically, on-demand and without post selection using a nucleus, which is controlled without entanglement with an ancilla, hence, is robust against error propagation, unlike continuous-variable systems~\cite{OPH+16Nature,GFP20Nature,HRO+17NatCommun,RRM+18Science,HWD19NatPhoton}.}
\textcolor{black}{Here, we consider SCSs (implying macroscopic to differentiate them from ``kitten" states~\cite{OTLG06Science}) 
on a nuclear spin, e.g., shown in Fig.~\ref{fig:nuclearcat}, but,
more generally, a superposition of any two maximally separated quasi-classical states, which are minimum uncertainty states that follow a classical evolution under a linear drive~\cite{p86springer,P72CommunMathPhys}.}
We are motivated by recent experiments on high-spin nuclei (spin $I>1$) in a solid-state device~\cite{AMJ+20Nature,dFBJ+23arxiv} accompanied by an enhanced electric field-driven quadrupole interaction, two orders of magnitude faster than the dephasing timescale, that could achieve ultrafast non-linear dynamics.

\textcolor{black}{
A highly coherent nuclear spin could open the possibility of robust generation and long-lived storage of SCSs.
So far, quadrupole interaction has only been utilized for realizing spin squeezed states (SQSs), that too, on an ensemble of nuclear spins~\cite{AAS+15PRL, KB16PRA}.
However, individual addressing of high-spin systems is important for applications like quantum information processing and error correction~\cite{G21PRL}.
We present two key results: i)~a scheme to go beyond squeezed states towards generating SCSs and ii)~an experimentally feasible control protocol for storing and manipulating SCSs on an individual nucleus.}


\section{Setup}

\begin{figure}
    \centering
    \includegraphics[width = \linewidth]{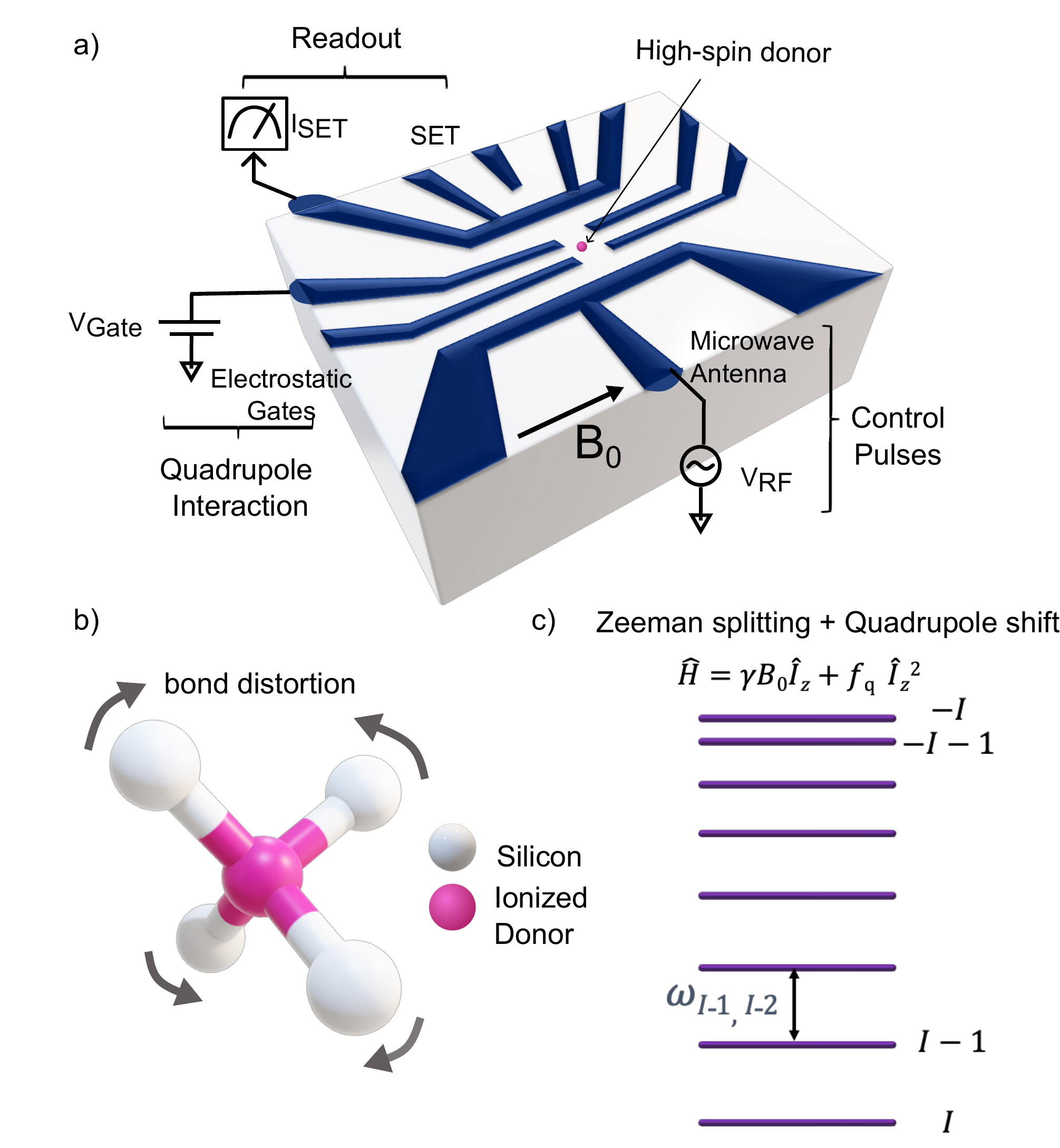}
    \caption{a) High-spin donor implanted in a fabricated silicon-based chip. b)~Bond distortion due to lattice strain leads to an electric-field gradient at the donor site. c) Energy-level diagram of a high-spin nucleus.}
    \label{fig:setup}
\end{figure}

We consider a high-spin donor (Group-V element with a high-spin nucleus) implanted in a nanoelectronic device based on enriched $^{28}$Si (or any other spinless Group-IV atoms), as described in~\cite{AMJ+20Nature}.
The chip is affixed to a dilution refrigerator maintained at a low temperature ($T\approx 0.01$~K) and placed inside a superconducting coil that produces a static magnetic field $\bm{B}_0 \approx 1$~T. 
As shown in Fig.~\ref{fig:setup}(a), the chip includes electrostatic gates to control the electrochemical potential of the donor, a microwave antenna to supply electromagnetic pulses, and a single electron transistor (SET) for reading out the donor spin state \cite{morello2020donor}. 

An implanted Group-V donor contains five electrons in its outermost shell, four of which form bonds with neighboring Si atoms and the remaining electron can be removed by ionizing the donor using electrostatic gates. 
The quantum state of the nuclear spin is initialized using flip-flop drive~\cite{savytskyy2023electrically}, where an electron and nuclear spin pair is flipped together in opposite directions. 
\textcolor{black}{The donor atom is subject to a static potential, and a target nuclear-spin eigenstate can be achieved by applying a combination of flip-flop pulses and electron spin-resonance pulses that move population from higher and lower nuclear spin states towards a target state \cite{dFBJ+23arxiv}.  }
Then, the electrostatic potential is increased to ionize the donor atom by removing the electron and only the nuclear spin is manipulated using RF pulses.~\cite{DPS+13Nano}. 
The readout process using electron spin resonance can be made dependent on the nuclear spin state, thus achieving nuclear readout via spin dependent tunneling of the electron~\cite{pla2013high,AMJ+20Nature}. 

The differential thermal expansion between the silicon substrate and the electrostatic gates induce lattice strain in the chip \cite{AMJ+20Nature}, thereby distorting the bonds between the donor atom and neighboring Si atoms, as shown in Fig.~\ref{fig:setup}(b). 
The electric-field gradient (EFG) generated from distortion of the bonds results in a quadrupole interaction of the high-spin nucleus, which has a non-zero quadrupole moment due to its asymmetric nuclear charge distribution. 
For a spatially varying potential $V(x,y,z)$, the EFG tensor comprises elements
$V_{\alpha\beta}
:= \frac{\partial^2 V(x,y,z)}{\partial\alpha\partial\beta},\,
\alpha,\beta \in \{x, y, z\},
$
which is real, traceless and symmetric.
Thus, $V_{\alpha\beta}$ can be diagonalized in a principal axis system $\{x'$, $y'$, $z'\}$, such that $|\mathcal{V}_{z'z'}|\geq |\mathcal{V}_{y'y'}| \geq |\mathcal{V}_{x'x'}|$, and the EFG is characterized by the asymmetry parameter $\eta := \frac{\mathcal{V}_{x'x'} - \mathcal{V}_{y'y'}}{\mathcal{V}_{z'z'}}$ for $0\leq \eta \leq 1$.
For a nucleus with electric quadrupole moment $q_\text{n}$, 
the quadrupole interaction is 
\begin{equation}
\label{eq:quadrupole}
    \hat{H}_\text{q} = \omega_{\text{q}} \left[  \hat{I}_{z'}^2 + \frac {\eta}{3}(\hat{I}_{x'}^2 - \hat{I}_{y'}^2 - \mathbb I^2)\right],
    \omega_\text{q} = \frac{3\text{e} q_\text{n} \mathcal{V}_{z'z'} }{4I(2I-1)\hbar},
\end{equation}
where~$\omega_\text{q}$ is the quadrupole coupling strength, 
e the elementary charge,
$\hbar$ the reduced Planck's constant, the nuclear spin operator vector $\hat {\bm {I}} := [\hat I_{x'}, \hat I_{y'},\hat I_{z'}]$ 
and invariant {$\mathbb I^2=\bm I \cdot \bm I$}.

In addition to the quadrupole interaction, we manipulate the  nuclear spin state using a static magnetic dipole interaction due to $\bm B_0$ and time dependent perturbation driven by magnetic field  $\bm{B}_1$ with time-dependence~$\epsilon(t)$ ($|\epsilon(t)|\leq1$). 
For a nucleus with gyromagnetic ratio $\gamma$, the nuclear spin dynamics are described by 
\begin{equation}\label{eq:hamiltonian}
    \hat H = \gamma \bm {\hat I} \cdot \bm B_0 + \hat H_{\text q} + \gamma \bm {\hat I} \cdot \bm B_1 \epsilon(t).
\end{equation}
The $2I+1$ degenerate levels of a nuclear spin undergo Zeeman splitting due to $\gamma \bm {\hat I} \cdot \bm B_0$ which creates an equal spacing between the energy levels. 
The electric quadrupole interaction $\hat H_{\text q}$ causes an unequal shift of the levels such that the transition frequency $\omega_{i,i-1}$  between two consecutive levels $i$ and $i-1$, for $i\in[I,-I+1]$, is unique, as shown in Fig.~\ref{fig:setup}(c). 
We represent spin states as $\ket{I,m_I}$, where $m_I \in [-I,I]$, and RF pulses drive transitions which follow the selection rule $\Delta m_I = \pm 1$.
For numerical simulations~\cite{JNN13CompPhysComm}, we use the example of a spin-$\nicefrac{7}{2}$ $^{123}$Sb donor~\cite{AMJ+20Nature,dFBJ+23arxiv}, with Zeeman splitting $\gamma B_0= 2\pi\times8.25$~MHz, $\omega_\text{q}=2\pi\times40$kHz and perturbation strength $\gamma B_1 = 2\pi\times800$~Hz.

\section{One-Axis Twisting}
\textcolor{black}{
Quadrupole interaction is equivalent to one-axis twisting (OAT) due to $\hat{I}^2_{z'}$ if $\eta=0$, two-axis counter twisting (TACT) due to $\hat{I}^2_{z'}$ and $\hat{I}^2_{y'}$ if $\eta=1$~\cite{KB16PRA,B17PRA}, and partial  TACT otherwise.
Generally, TACT is preferred over OAT as TACT generates faster squeezing in comparison to OAT and can help produce SQSs for metrology~\cite{LXJY11PRL}. 
However, TACT cannot produce SCSs, and, in general, quadrupolar non-linear evolution with $\eta>0$ cannot go beyond SQSs on a nuclear spin~\cite{AAS+15PRL}. 
}

\textcolor{black}{We devise a way to convert TACT to OAT by adding a large
linear term to the quadrupole interaction, induced here by the Zeeman coupling.
The resulting dynamics depend on the orientation of the EFG and strength of magnetic field $B_0$.} 
When the Zeeman splitting is much larger than quadrupole coupling, i.e.~$\gamma\bm B_0\gg \omega_\text{q}$, the component perpendicular to $\bm B_0$ does not affect the nuclear spin because quadrupolar transitions ($\Delta m_I =\pm 2$) are prevented by the large energy difference.
In this regime, only the component of the quadrupole interaction parallel to $\bm B_0$ generates non-linear rotation, which effectively results in OAT, regardless of the value of $\eta$.
In contrast to TACT, our scheme allows generation of SCSs for any orientation of the quadrupolar axes of symmetry $\{x',y',z'\}$ with respect to linear interaction along the $z$ axis. 
For numerical simulations in this work, we set $\bm B_0 = B_0 \hat z$, $\bm B_1 = B_1 \hat y$, $\eta=0$, and  EFG to be symmetric about $z$ axis, i.e.~$\hat z'=\hat z$, but, our results can be generalized to other configurations. 
\begin{figure}
    \centering
    \includegraphics[width = \linewidth]{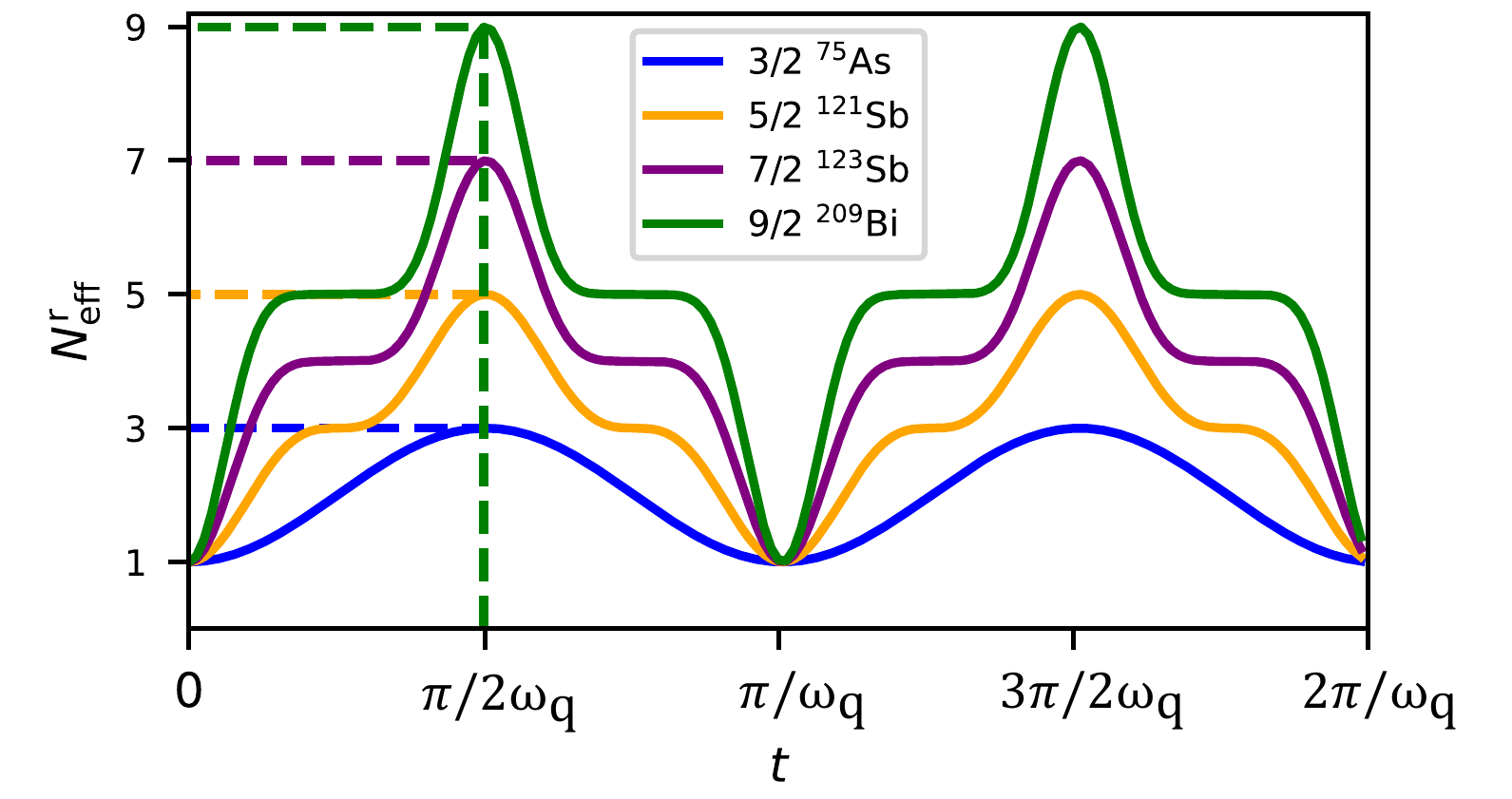}
    \caption{Degree of superposition $N_{\text{eff}}^\text{r}$ under free evolution (\textcolor{black}{$\epsilon(t)=0$}) of a coherent spin state to a SCS at $t=\nicefrac{\pi}{2\omega_\text q}$.
    For differing nuclear spins, the maximum value $N_{\text{eff}}^\text{r}\propto2I$, showing the macroscopicity of these SCSs\textcolor{black}{, but the $I=\nicefrac32$ case lacks a ``hill-on-mesa'' and is actually a ``spin `kitten' state''.}}
    \label{fig:twisting}
\end{figure}

Despite the femtometer scale of a nucleus, nuclear-SCSs are macroscopic with their size,  quantified by the relative quantum Fisher information~\cite{FD12NJP,FSD+18RevModPhys}, scaling linearly with $I$.
The size or degree of superposition $N_{\text{eff}}^\text{r}$ of a nuclear-spin state~$\psi$ for measurement operator $\hat O$ is
\begin{equation}
\label{eq:catness}
    N_{\text{eff}}^\text{r}
    =\frac2I
    \left(\bra{\psi}\hat O^2\ket{\psi}-\bra{\psi}\hat O\ket{\psi}^2\right),
\end{equation}
i.e., proportional to the variance of the observable.
For an appropriate $\hat O$, the variance is $I^2$ for bimodal SCSs and we calculate the size of a nuclear SCS to be $2I$, which denotes the maximum quantum Fisher information of the system and is equivalent to a GHZ state~\cite{GHZ89} made of $N=2I$ entangled qubits~\cite{CBE+18NatComm}. 
A nuclear SCS with $2I+1$ dimensions, e.g., a spin-$\nicefrac{7}{2}$ SCS with 8 dimensions is equivalent to a $2I=7$ qubit GHZ state with permutation symmetry, and not a $\log_2 (2I+1) = 3$ qubit  state, and can have macroscopicity up to $2I=9$ for a $^{209}$Bi donor.

The degree of superposition $N_{\text{eff}}^\text{r}$ (Eq.~\eqref{eq:catness}) increases and decreases periodically during OAT, as shown in Fig.~\ref{fig:twisting}.
$N_{\text{eff}}^\text{r}=1$ for a spin coherent state \textcolor{black}{on the $\hat I_x$ axis of the Bloch sphere} at $t=0$, increases as the spin state squeezes, then reaches the flat-top of the ``mesa" where it interferes with itself and finally increases again (``hill") as two distinct components form resulting in a SCS at $t=\nicefrac{\pi}{2\omega_\text{q}}$.
Under further evolution, the SCS returns to a coherent state at $t=\nicefrac{\pi}{\omega_\text{q}}$, which again squeezes to a SCS at $t=\nicefrac{3\pi}{2\omega_\text{q}}$, and so on.
This periodic process of collapse-and-revival of the SCS has a time period of $\nicefrac{\pi}{\omega_{\text{q}}}$ and is a signature of coherence~\cite{CBE+18NatComm}.
From the plots for  different values of nuclear spin in Fig.~\ref{fig:twisting}, we note the macroscopicity of the SCSs through the linear scaling of their size $(N^\text{r}_{\text{eff}})_{\text{max}}=2I$.

\section{Control Scheme}
\textcolor{black}{External control of a high-spin nucleus is required for initialization, manipulation and detection of nuclear-SCSs. }
For initialization, \textcolor{black}{the eigenstate $\ket{I,I}$, which is also a coherent state}, should first be rotated to reside on the $\hat I_x$ axis of the Bloch sphere. 
Then, OAT results in a SCS on the \textcolor{black}{$\hat O= \hat I_y$} axis for a half-integer nuclear spin, but, for donors in solid-state devices, nuclear-spin initialization and measurement is limited to the eigenstates of $\hat I_z$. 
\textcolor{black}{Observing nuclear SCSs could be possible with angular momentum operations, which are trivial when~$\gamma B_1 \gg \omega_\text{q}$~\cite{B17PRA,OMMD21PRA} or  with a tunable non-linearity~\cite{PSO+18RMP}.
However, quantum control is needed as, experimentally, $\gamma B_1 = 2\pi\times 800$~Hz $\ll \omega_\text{q}=2\pi\times40$~kHz such that a RF pulse can only selectively drive a particular nuclear spin transition leading to Givens rotations, and the quadrupole coupling cannot be turned on-and-off.}

For applying global SU(2) rotations with an always-on non-linearity and low drive strength, we use a multi-tone (MT) pulse that  is a superposition of $2I$ tones of equal amplitudes, corresponding to the $2I$ nearest neighbor sub-spaces with transition frequencies represented by frequency vector $\bm\omega:=\{\omega_{i,i-1}\}_{i=I}^{-I-1}$.
Each tone has a global phase $\phi$ and a pulse envelope switched on at a time $t_0$ is described by the amplitude $\epsilon^{\text{MT}}(t;\bm\omega,\phi,t_0) =  \frac{ 1}{2I}\sum_{\omega_{j}\in\bm\omega}\cos(\omega_{j} (t-t_0) + \phi)$.
Here, $\phi$ tunes the axis of rotation on the Bloch sphere while preserving the relative phase between different energy levels.
The frequency $\Omega$ of SU(2) rotations is generalized from the Rabi frequency $\nicefrac{\gamma B_1}{2}$ for a two-level nuclear spin, to a high-spin nucleus by dividing the Rabi frequency by the number of tones $2I$, i.e.~$\Omega = \nicefrac{\gamma B_1 }{ 4I}$.
Applying this pulse for time $\Delta t$ rotates the nuclear spin by angle $\Theta$ given by $\Theta = \Omega \Delta t$.
\textcolor{black}{In contrast to other schemes that employ numerical techniques with multi-tone control~\cite{LL03PRB,GFB+17PRL}, we derive exact expressions for phase-locked pulses, 
which can enable novel experimental applications such as full-state tomography and interferometry with a high-spin nucleus.}

\begin{figure*}
    \centering
    \includegraphics[width = \linewidth]{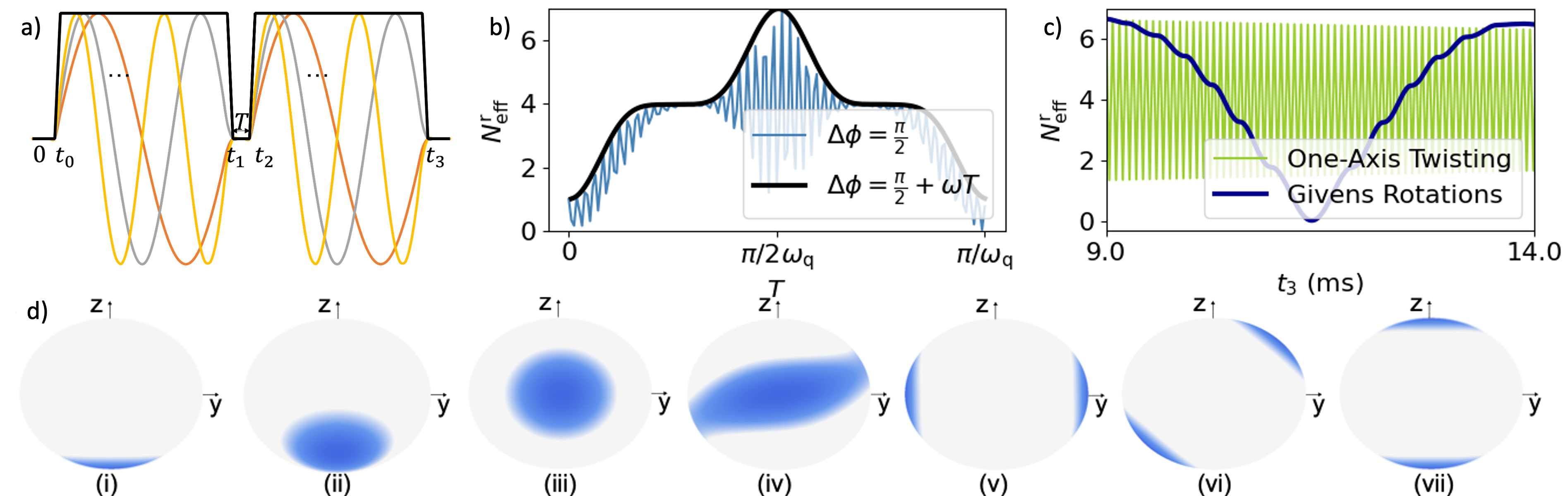}
    \caption{a) Generalized Ramsey-like protocol for creating a SCS (Eq.~\eqref{eq:scheme2pulse}) 
    using multi-tone pulses. 
    b) Observing SCS collapses-and-revivals by varying $T$, where $\Delta\phi = \nicefrac{\pi}{2}+\omega T$ (black) is used for rotating-frame transformation. c) Collapses-and-revivals under decoherence with \textcolor{black}{$\Gamma_\text{m} = 10$~Hz and $\Gamma_\text{e} = 0.1$~Hz}. d) Dynamics illustrated by Husimi Q-function plots.}
    \label{fig:control}
\end{figure*}

Figure \ref{fig:control}(a) shows our Ramsey-like control sequence to realize SCSs, where a square-shaped MT RF pulse is applied \textcolor{black}{to the initial state $\ket{I,I}$} between time $t_0=0$ and $t_1 = t_{\nicefrac \pi 2}$. Then the pulse is switched off for time $T$, and again applied between $t_2=T+t_{\nicefrac \pi 2}$ and $t_3=T+2t_{\nicefrac \pi 2}$, with the amplitude 
\begin{align}
\label{eq:scheme2pulse}
\epsilon^{\text{cat}}(t;\bm\omega,\Delta\phi,T)
=&\epsilon^{\text{MT}}(t;\bm\omega,0,t_0)\sqcap\left(\textcolor{black}{t};t_0,t_1\right)\nonumber\\
&+\epsilon^{\text{MT}}(t;\bm\omega,\Delta\phi,t_2)\sqcap\left(\textcolor{black}{t};t_2,t_3\right)\big]
\end{align}
for~$\sqcap$ the top-hat function.
Here, the global rotation of $\Theta = \nicefrac \pi 2$ would take time $\Delta t_{\nicefrac \pi 2} = \frac{\pi}{2}\cdot\frac{4I}{\gamma B_1}$ and setting the phase difference $\Delta\phi=\nicefrac{\pi}{2}$ between the first and the second global rotations accounts for the angle between a coherent state and the resulting SCS on the equator.
The resulting dynamics are shown in Fig.~\ref{fig:control}(d), where we note that SCSs along any axis can be realized using global rotations (Fig.~\ref{fig:control}(d)(vi)).
The quadrupole coupling cannot be turned off and these states squeeze during rotations. 
A SCS can be stored for long times by aligning it to the fixed points on the $z$ axis (Fig.~\ref{fig:control}(d)(vii)).

For a spin-$\nicefrac{7}{2}$ $^{123}$Sb nucleus, Fig.~\ref{fig:control}(b) shows the change in $N_{\text{eff}}^\text{r}$,
for $\hat O = \hat I_z$,
as $T$ is varied, similar to Fig.~\ref{fig:twisting}.
When $\Delta\phi = \nicefrac{\pi}{2}$, we observe a rapid variation of $N_{\text{eff}}^\text{r}$ with a frequency of $\nicefrac{\gamma B_0}{\pi}$, because nuclear-spin states generated by OAT undergo Larmor precession and when the axis of rotation corresponds with the axis of the SCS, the second global rotation is unable to move the state.
To remedy this immobility, we assign a $T$-dependent shift to the phase difference: $\Delta \phi = \nicefrac{\pi}{2}+\omega T$ in~(\ref{eq:scheme2pulse}), which is equivalent to a rotating-frame transformation, and results in the periodic collapses-and-revivals of a SCS solely due to OAT, which would enable efficient experimental detection.

\section{Virtual Phase Control}
\textcolor{black}{In this section, we show that cat-state generation is possible with just phase modulation of multi-tone pulses in the generalized rotating frame of reference. First, we note that,
in the limit of large Zeeman interaction $\gamma B_0\gg\omega_{\text q}$,
we can neglect non-commuting terms, such as~$\hat I_{x'}^2$ of the quadrupole interaction~(\ref{eq:quadrupole}) so that the non-linear evolution is only described in terms of the strength of quadrupole tensor, $\omega_{\text{q}}^{\text{eff}}$, parallel to the magnetic field $\bm B_0$.
Under this approximation and using the notation $[\ell,\ell']:=\{\ell,\ell+1,\dots,\ell'\}$,
the effective Hamiltonian is
\begin{equation}
    \hat H_{\text{eff}} = \gamma B_0\hat I_z + \omega_{\text{q}}^{\text{eff}} \hat I_z^2 + \gamma B_1 \hat I_x\sum_{j\in[1,2I]} \epsilon_j \cos(\omega_{j}t+\phi_j),
\end{equation}
where $\omega_j$, $\epsilon_j$, and $\phi_j$ are respectively the frequency, amplitude and phase of the $j^{\text{th}}$ component of the multi-tone control. The eigenstates $\ket{I,k}$ of the nuclear spin have energy
\begin{equation}
e_k = \gamma B_0 k + \omega_\text{q}^{\text{eff}}k^2
\end{equation}
for $k\in [-I,I]$.}

\textcolor{black}{Now we transform the Hamiltonian to the generalized rotating frame, defined by the diagonal matrix~\cite{LL03PRB}
\begin{equation}
U_{\text{rot}} = \text{diag}\left( e^{-ie_It}, e^{-ie_{I-1}t}, \ldots \right)
\end{equation}
and obtain 
\begin{equation}
    \hat H_{\text{rot}} = \gamma B_1 \begin{bmatrix}
        0 & h_1\epsilon_1 e^{-i\phi_1} & 0 & \ldots\\
        h_1\epsilon_1 e^{i\phi_1} & 0 & h_2\epsilon_2 e^{-i\phi_2} & \ldots\\
        0 & h_2\epsilon_2 e^{i\phi_2} & 0 & \ldots \\
        \vdots & \vdots & \vdots & \ddots
    \end{bmatrix},
\end{equation}
where
\begin{equation}
h_j = \bra{I,I-(j-1)}\hat I_x\ket{I,I-j}.
\end{equation}
We note that global rotations are applied by setting $\epsilon_j=\nicefrac{1}{2I}$ and $\phi_j=\phi$; for example, using $\phi=0$ is equivalent to
\begin{equation}
\hat H_{\text{rot}} = \frac{\gamma B_1\hat I_x}{2I}
\end{equation}
and $\phi=\nicefrac{\pi}{2}$ is equivalent to $\hat H_{\text{rot}} = \nicefrac{\gamma B_1\hat I_y}{2I} $.}

\textcolor{black}{
One-axis twisting can be realized by virtually updating the phases of a multi-tone pulse. To see this, we note that that the effective quadrupolar operator applied for duration $T$ generates the operation $e^{-iT\omega_{\text{q}}^{\text{eff}}\hat I_z^2 }$, which results in  a non-linear phase shift between the different eigenstates of a nuclear-spin Hamiltonian.
Using the dummy variable~$k$ for integers in the interval~$[-I,I]$,
a state
\begin{equation}
\ket \psi = \sum_{k\in[-I,I]} c_k \ket{I,k}
\end{equation}
is transformed as
\begin{equation}
    e^{-iT\omega_{\text{q}}^{\text{eff}}\hat I_z^2 }\ket \psi = \sum_{k\in[-I,I]} c_k e^{-iT\omega_{\text{q}}^{\text{eff}}k^2} \ket{I,k},
\end{equation}
where $-T\omega_{\text{q}}^{\text{eff}}k^2$ is the resulting phase shift on the energy eigenstate $\ket{I,k}$.}

\textcolor{black}{%
Now, we can factor out the global phase $-T\omega_{\text{q}}^{\text{eff}}I^2$ and
instead of freely evolving a state under one-axis twisting, apply equivalent phase updates,
\begin{equation}
\phi_j\mapsto\phi_j+T\omega_{\text{q}}^{\text{eff}}\left((I-j)^2-I^2\right),
\end{equation}to the subsequent multi-tone pulse. 
Such virtual phase updates would have same effect as free non-linear evolution and would result in cat state formation when $T=\nicefrac{\pi}{2\omega_{\text{q}^\text{eff}}}$. 
Thus, in the generalized rotating frame, we can start with the initial state $\ket{I,I}$, apply a global $\nicefrac{\pi}{2}$ rotation with $\phi_j = \phi$, followed by a second by a global $\nicefrac{\pi}{2}$ rotation with
\begin{equation}
\phi_j = \phi+\nicefrac{\pi\left((I-j)^2-I^2\right)}{2},
\end{equation} 
realizing cat states by only phase modulation of multi-tone pulses.}

\section{Decoherence}
Cat states with a lifetime of several milliseconds can be achieved using silicon-based chips with an implanted spin-$\nicefrac{7}{2}$ $^{123}$Sb donor atom, that have a nuclear spin coherence time $T_2^* \approx 100$~ms for two-level transitions inferred from experiments on nuclear electric resonance~\cite{AMJ+20Nature}.
SCSs along the $z$-axis would have a reduced coherence time, lower by a factor of $2I$,  due to the large separation between the two components, 
resulting in a coherence time of $\nicefrac{100}{7} \approx 14$~ms for $^{123}$Sb. 
Even for a modest driving strength $\gamma B_1=2\pi\times800$~Hz, a $\nicefrac{\pi}{2}$-global rotation would take $\Delta t_{\nicefrac \pi 2} = \frac{\pi}{2}\cdot\frac{4I}{\gamma B_1} = 4.375$~ms.
Thus, our approach, using two MT pulses and OAT, can yield highly coherent SCSs in less than 9~ms---within the dephasing time on current hardware.
Faster Rabi frequencies are easily achieved for nuclear magnetic resonance (NMR) methods~\cite{pla2013high,dFBJ+23arxiv}, that could reduce the time needed for MT pulses and lead to faster generation and detection of nuclear SCSs.

We compare the performance of OAT  to an intuitively appealing gate sequence for creating SCSs, achieved by sequential Givens rotations~\cite{C01CSE}.
Starting with~$\ket{I, I}$,
we apply a $\nicefrac{\pi}{2}$-pulse of frequency $\omega_{I,I-1}$ to form the superposition $\ket{I,I}+\ket{I,I-1}$,
with unit norm implied. 
Then, we apply a sequence of~$2I-1$ $\pi$-pulses to create the SCS $\ket{I,I}+\ket{I,-I}$. 
For observing a collapse of the SCS, repeating 
the same sequence
evolves the SCS into the coherent state $\ket{I,-I}$, opposite to the initial state $\ket{I,I}$.
In total, $4I$ pulses would be needed for each collapse-and-revival using Givens rotations, that would take $\sim9$~ms for a $I=\nicefrac{7}{2}$ with Rabi frequency  $2\pi\times800$~Hz, in contrast to OAT which takes just $\nicefrac{\pi}{\omega_\text q}=12.5\mu$s when $\omega_\text{q}=2\pi\times40$~kHz.

We simulate our scheme under decoherence from a magnetic field fluctuating at rate $\Gamma_\text{m}$, modeled using Lindblad operator \textcolor{black}{$\hat L_\text{m} = \hat{I}_z$}, and an electric field fluctuating at rate $\Gamma_\text{e}$, modeled using \textcolor{black}{$\hat L_\text{e} = \hat{I}_z^2$}. 
Figure~\ref{fig:control}(c) shows $N_\text{eff}^{r}$ vs evolution time beyond 9~ms (approximate time to form the first SCS).
\textcolor{black}{We note that OAT is two orders of magnitude faster than both Givens rotations and the dephasing timescale, opening the possibility of realizing hundreds of SCS collapses-and-revivals
by using a strong non-linearity relative to decoherence, which makes our SCS robust,
in contrast to alternative platforms~\cite{KVL+13Nature}. \textcolor{black}{Additionally, SCSs generated by Givens rotations lie on the $\hat I_z$ axis, where the dephasing rate of the cat state increases with the nuclear spin increases; in comparison, cat states produced by one-axis twisting lie on the equator, perpendicular to the dephasing axis, and are highly protected against decoherence.}}

\section{Conclusion}

\textcolor{black}{
In summary, 
Schr\"odinger cat states are important for investigating foundations of quantum mechanics as well as for applications such as quantum error correction but are fragile and hard to realize.
We propose a new platform for realizing macroscopic spin cat states by converting quadrupolar two-axis counter twisting to one-axis twisting on a high-spin donor in a solid-state device. 
Our scheme can  deterministically generate spin cat states on-demand without post-selection and is two orders of magnitude faster than the dephasing timescale, resulting in highly coherent states. 
\textcolor{black}{We show that, in the generalized rotating frame, a spin coherent state can be converted to a cat states by only phase modulation of multi-tone pulses.}
Unlike continuous-variable systems, we generate cat states without entanglement with an ancilla, which renders robustness against error propagation~\cite{RRM+18Science} and opens the possibility of storing long-lived nuclear-spin cat states.
Furthermore, our cat states can be experimentally realized on a single nucleus, which could lead to novel applications in quantum information processing.}

\begin{acknowledgments}
 
We thank Dipankar Home for useful discussion on macroscopicity of nuclear cat states. P.G.~and B.C.S.~acknowledge support from NSERC. 
Work at UNSW was funded by an Australian Research Council (DP210103769). A.V.~and X.Y.~acknowledge support from the Sydney Quantum Academy. 

\end{acknowledgments}
\bibliography{main}

\appendix
\section{Dynamics}
\subsection{Quadrupole interaction}
Our system is described by the Hamiltonian 
\begin{equation}\label{eq:SMHamiltonian}
    \hat H = \gamma B_0 \hat I_z + \sum_{\alpha,\beta} Q_{\alpha\beta}\hat I_\alpha \hat I_\beta + \gamma B_1 \hat I_y \epsilon(t),
\end{equation}
where $\gamma$ is the nuclear gyromagnetic ratio, $B_0$ is the strength of the static magnetic field oriented along $z$ axis, $B_1$ is the strength of perturbative magnetic field and $\epsilon(t)$ the time-dependent pulse amplitude.
The quadrupole moment tensor components are
\begin{equation}
\label{eq:Qaphabeta}
    Q_{\alpha\beta} = \frac{3\text{e} q_\text{n} \mathcal{V}_{\alpha\beta} }{4I(2I-1)\hbar},
\end{equation}
with~$q_\text{n}$ the electric quadrupole moment, which characterizes the spherical asymmetry of the nuclear charge distribution, e the elementary charge,
$\hbar$ the reduced Planck's constant, and
$V_{\alpha\beta}
:= \frac{\partial^2 V(x,y,z)}{\partial\alpha\partial\beta}$ the electric field gradient.
The quadrupole moment tensor~(\eqref{eq:Qaphabeta}) is real, traceless and symmetric; thus, a set of principal axes of symmetry (PAS) $\{x', y', z'\}$ exist for which the quadrupole moment tensor can be diagonalized.

The quadrupole term in the Hamiltonian can be expressed in the PAS as
\begin{equation}
\label{eq:SMquadrupole}
    \hat{H}_\text{q} = \omega_{\text{q}} \left[  \hat{I}_{z'}^2 + \frac {\eta}{3}(\hat{I}_{x'}^2 - \hat{I}_{y'}^2 - \mathbb I^2)\right],
\end{equation}
where $\eta := \frac{\mathcal{V}_{x'x'} - \mathcal{V}_{y'y'}}{\mathcal{V}_{z'z'}}$ is the asymmetry parameter for $0\leq \eta \leq 1$, and  $\mathbb{I}^2=\bm I \cdot \bm I$.
The orientation of the PAS with respect to the lab frame of reference can be expressed using Euler angles $\{\delta,\mu,\nu\}$ as
\begin{equation}\label{eq:SMeulerx}
\begin{split}
    \hat I_{x'} = &\left( \cos \nu \cos \delta - \sin \nu \cos \mu \sin \delta \right) \hat I_x + \\
     &\left( \cos \nu \sin \delta - \sin \nu \cos \mu \cos \delta \right) \hat I_y + \\
     &\sin\nu\sin\mu \hat I_z,
\end{split}    
\end{equation}
\begin{equation}\label{eq:SMeulery}
    \begin{split}
       \hat I_{y'} = &\left( -\sin \nu \cos \delta + \cos \nu \cos \mu \sin \delta \right) \hat I_x + \\
        &\left( -\sin \nu \sin \delta + \cos \nu \cos \mu \cos \delta \right) \hat I_y + \\
        &\cos\nu\sin\mu \hat I_z,
    \end{split}
\end{equation}
\begin{equation}\label{eq:SMeulerz}
    \hat I_{z'} = \sin \mu \sin\delta \hat I_x -\sin \mu \cos\delta I_y + \cos \mu I_z.
\end{equation}
Together with $\eta$, the three Euler angles completely specify the symmetric and traceless quadrupole tensor for a given nucleus.

The EFG depends on several factors including microscopic interactions within the crystal lattice, the position of the donor, the magnitude of the applied electric field and the fabrication process of the device.
In general, it is hard to pre-determine the EFG orientation or quadrupole splitting; however, finite-elements modelling can be used to estimate the strength of quadrupole interaction. 
In recent experiments~\cite{AMJ+20Nature,dFBJ+23arxiv}, a quadrupole splitting of up to $2\pi\times 60$~kHz was observed.

\subsection{Macroscopicity}
\begin{figure}[t]
    \centering
    \includegraphics[width = \linewidth]{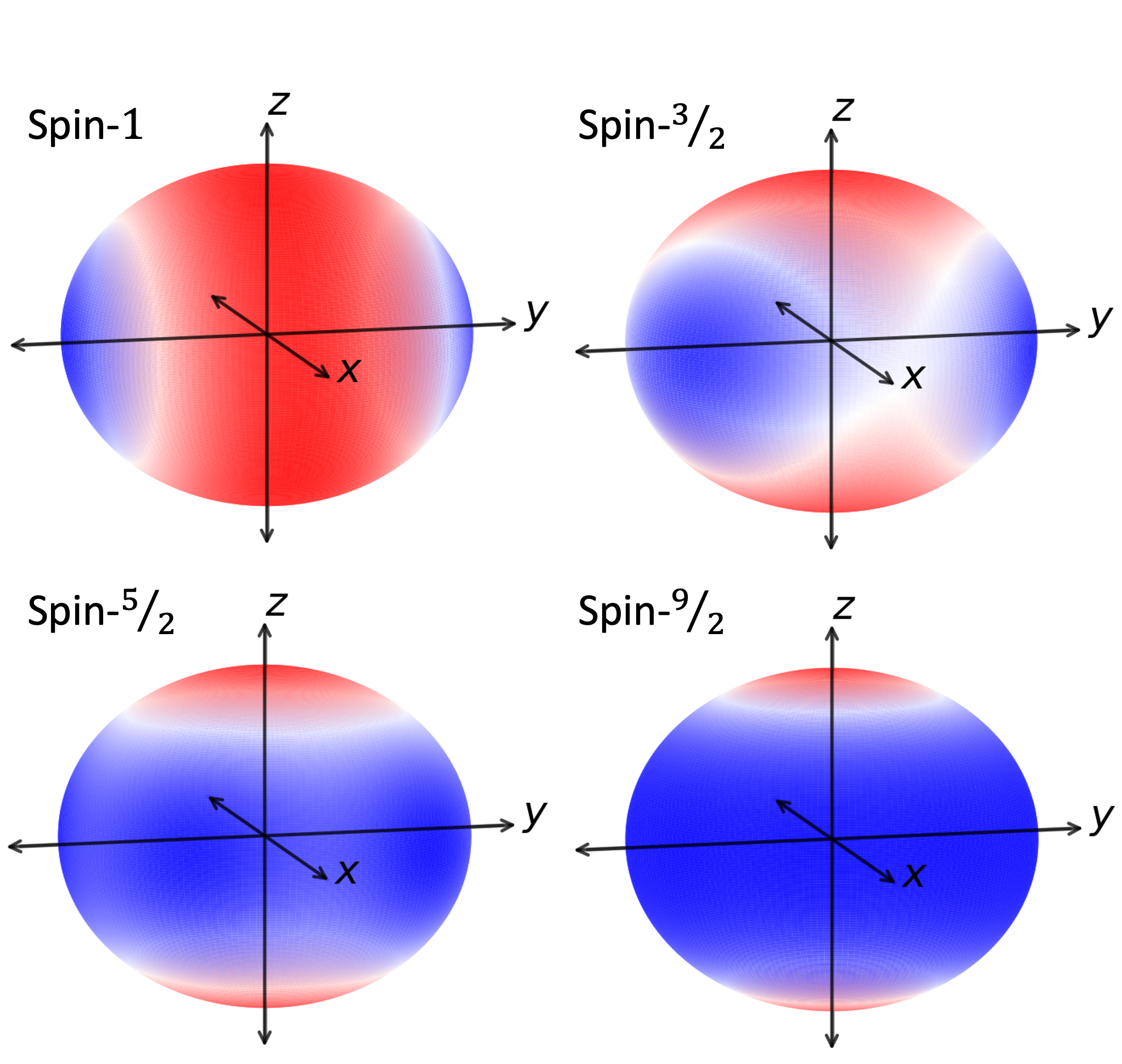}
    \caption{Husimi Q-function plots of  SCSs for different values of the spin. The spin-$1$ case lacks bi-modality as the spin coherent states overlap with each other. The spin-$\nicefrac{3}{2}$ case corresponds to a ``spin `kitten' state'' with small overlap between its two components. Higher spin ($\geq\nicefrac{5}{2}$) form macroscopic cat states.}
    \label{fig:kittenandcat}
\end{figure}

Figure~\ref{fig:kittenandcat} shows the Husimi Q-function plots of superpositions of maximally separated coherent states along the $\hat{I}_z$ axis of the Bloch sphere. We note that spin-$1$ case lacks bi-modality and cannot form cat states. Spin-$\nicefrac{3}{2}$ case forms spin kitten states and higher spins ($\geq\nicefrac{5}{2}$) are required for forming cat states (implying macroscopic).

\section{Experimental Control}\label{SMsec:experiment}

\subsection{Initialization and readout}\label{sec:init_readout}
\begin{figure}
    \centering
    \includegraphics[width = \linewidth]{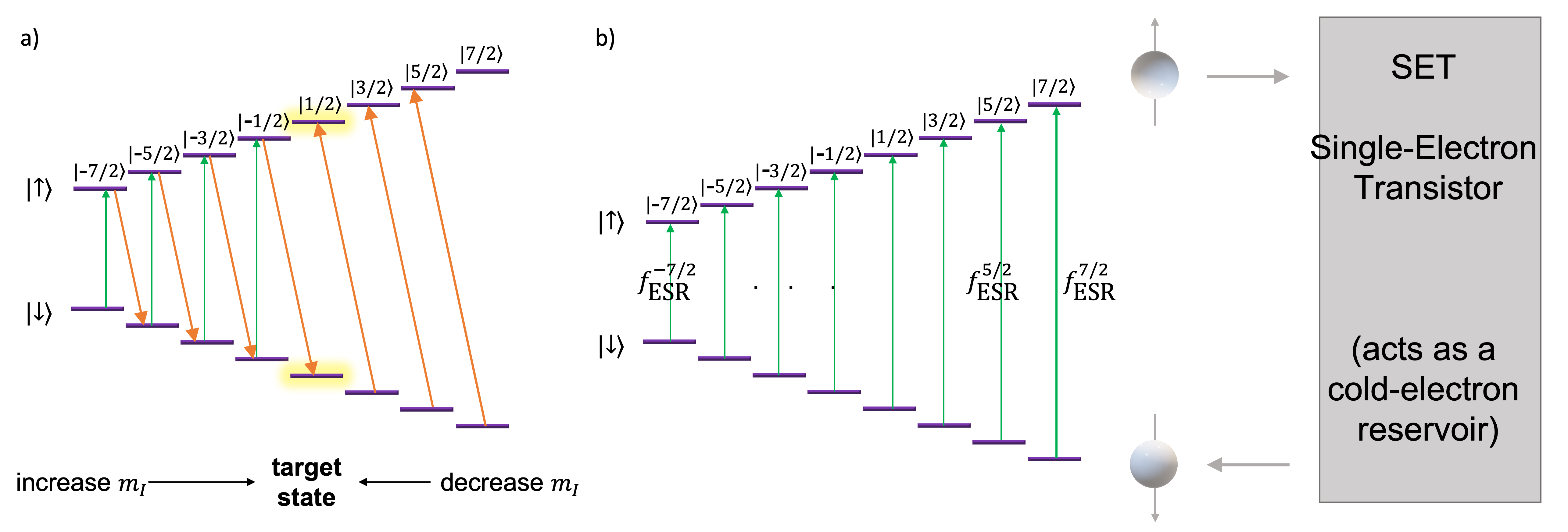}
    \caption{Initialization (a) and readout (b) of a nuclear spin by coupling it to an electron spin and using flip-flop pulses (orange arrows) and  electron-spin resonance pulses (green arrows)}.
    \label{fig:flipflop}
\end{figure}

The nuclear spin of the donor atom can be coherently controlled through electromagnetic interaction using the electrostatic gates and microwave antenna fabricated on the chip.
The donor atom is subjected to a static potential, with the nuclear spin coupled to an electron spin, and initialization of the nuclear spin is done via electron-nuclear hyperfine coupling. 
A typical experiment uses flip-flop driving~\cite{savytskyy2023electrically}, in which an electron spin is flipped and a nuclear spin is flopped in opposite directions and high-power chirp pulses, corresponding to different transitions, are applied sequentially. 
As shown in Fig.~\ref{fig:flipflop}~(a),
to reach a target state in the energy ladder, first, the nuclear spin is pumped down from all higher-lying energy levels using flip-flop transitions (orange). 
Then, the nuclear spin state is pumped up from all the lower-lying energy levels using a combination electron-spin resonance pulses (green) and flip-flop pulses (orange).

The initialization and readout schemes benefit from the coupling of a nuclear spin to an electron spin, which enables fast initialization of the nuclear spin to any energy eigenstate $\ket{I,m_I}$, of which $\ket{I,\pm I}$ are spin coherent states. We use the ground state $\ket{I,I}$ as the initial state in our scheme.
After initialization, the electrostatic potential is increased to ionize the donor atom by removing the electron and only the nuclear spin is manipulated using RF pulses~\cite{DPS+13Nano}. 
When the nuclear spin reaches the desired final state, the donor atom is de-ionized such that the nuclear spin again couples to an electron. 
As shown in Fig.~\ref{fig:flipflop}~(b), the nuclear-spin state is read out using single-shot detection with pulsed electron spin resonance and spin-dependent tunneling of the electron to the SET, which acts as a cold-electron reservoir~\cite{MPZ+10Nature}.

\subsection{Control}
We manipulate the nucleus using RF magnetic pulses that cause NMR transitions with selection rule $\Delta m_I = \pm 1$. The quadrupole interaction shifts the energy levels of a nuclear spin qudit, as shown in Fig. 2 of the main text, which allows selectively addressed transitions by tuning the frequency of Rabi pulses, applied through the time-dependent perturbation $\epsilon(t)$ in Eq.~\eqref{eq:SMHamiltonian}. 
At off-resonant drives, the probability of a Rabi flip is given by,
\begin{equation}
    \text{P}_\text{flip} = \frac {(\gamma B_1)^2} {\Omega^2} \sin^2 \Omega t,
\end{equation}
where $\gamma B_1$ is the drive strength~\eqref{eq:SMHamiltonian}, $\Omega^2 = \sqrt{(\gamma B_1)^2 + \Delta \omega^2/4}$  with $\Delta \omega$ the difference between the detuning of the drive from the transition frequency.
With quadrupole interaction $\omega_\text{q}\approx 2\pi\times 40$~kHz achieved in experiments~\cite{AMJ+20Nature,dFBJ+23arxiv} and drive strength $\gamma B_1 = 2\pi\times0.8$~kHz, frequency difference 
between two transitions $\Delta\omega \geq 2\omega_\text{q}$, thus, $\text{P}_\text{flip}<0.0004$.
Thus, cross talks are highly suppressed.
Additionally, the quadrupole interaction generates fast collapse and revivals due to non-linear evolution having time-period of a few microseconds, which can be resolved by pulses applied using an Arbitrary Waveform Generator (AWG) with a resolution of 1~ns.
While lower-order non-linearity would slow down SCS generation, higher-order non-linearity, for example in the GHz range, would generate OAT at a rate faster than the resolution of an AWG.

\section{Decoherence}

\subsection{Master equation for a high-spin nucleus}
We model decoherence using a Lindblad master equation of the form
\begin{equation}\label{eq:SMmastereq}
\begin{split}
    \Dot \rho = -i [\hat H,\rho]  
    & + \Gamma_\text m \left(\hat L_\text m\rho \hat L_\text m^\dagger - \frac 1 2\{\hat L_\text m^\dagger \hat L_\text m,\rho\} \right) \\
    & + \Gamma_\text e \left(\hat L_\text e\rho \hat L_\text e^\dagger - \frac 1 2\{\hat L_\text e^\dagger \hat L_\text e,\rho\} \right),
\end{split}
\end{equation}
where $\rho$ is the density matrix of the nuclear spin, $\Gamma_\text m$ is the decay rate  due to magnetic field fluctuations, modeled using Lindblad operator $\hat L_\text m = \hat I_z$ and $\Gamma_\text e$ is the decay rate due to electric field fluctuations, modeled using Lindblad operator $\hat L_\text e = \hat I_z^2$. Here, we take the electric field fluctuation only in the $z$ direction, i.e.~parallel to the Zeeman splitting, because other components of the electric quadrupole interaction are washed out, as described above.

\begin{figure}
    \centering
    \includegraphics[width = \linewidth]{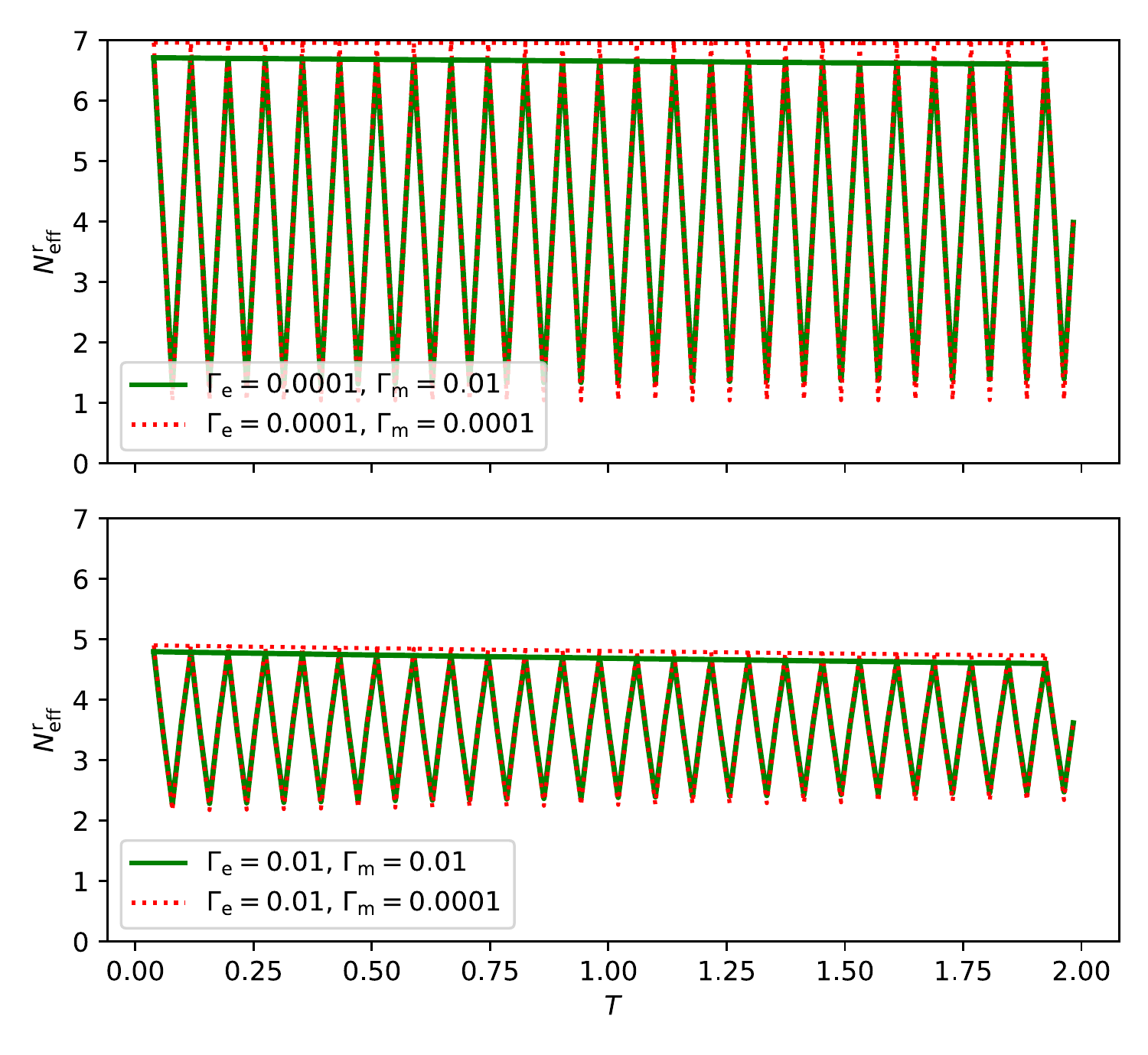}
    \caption{Collapse and revivals under decoherence~\eqref{eq:SMmastereq}. The line joining the revival points (where $N^\text{r}_{\text{eff}}$ is maximum) can be extrapolated to estimate the number of revivals possible within the coherence time of a nuclear spin. The decoherence rates are specified in units of kHz and time in ms.}
    \label{fig:decoherence}
\end{figure}

We simulate the cat-state collapse-and-revival control scheme for various values of $\Gamma_{m}$ and $\Gamma_{e}$, shown in Fig.~\ref{fig:decoherence}.
Here, magnetic fluctuations slowly dephase the nuclear spin and the coherence of the SCS decreases during collapse-and-revivals.
On the other hand, electric field fluctuations have a more drastic effect on SCSs generated by quadrupole interaction, resulting in an overall decrease of the variance, i.e.~macroscopicity of the SCSs.
\subsection{Coherence time of a qudit SCS}
\begin{figure}
    \centering
    \includegraphics[width  = \linewidth]{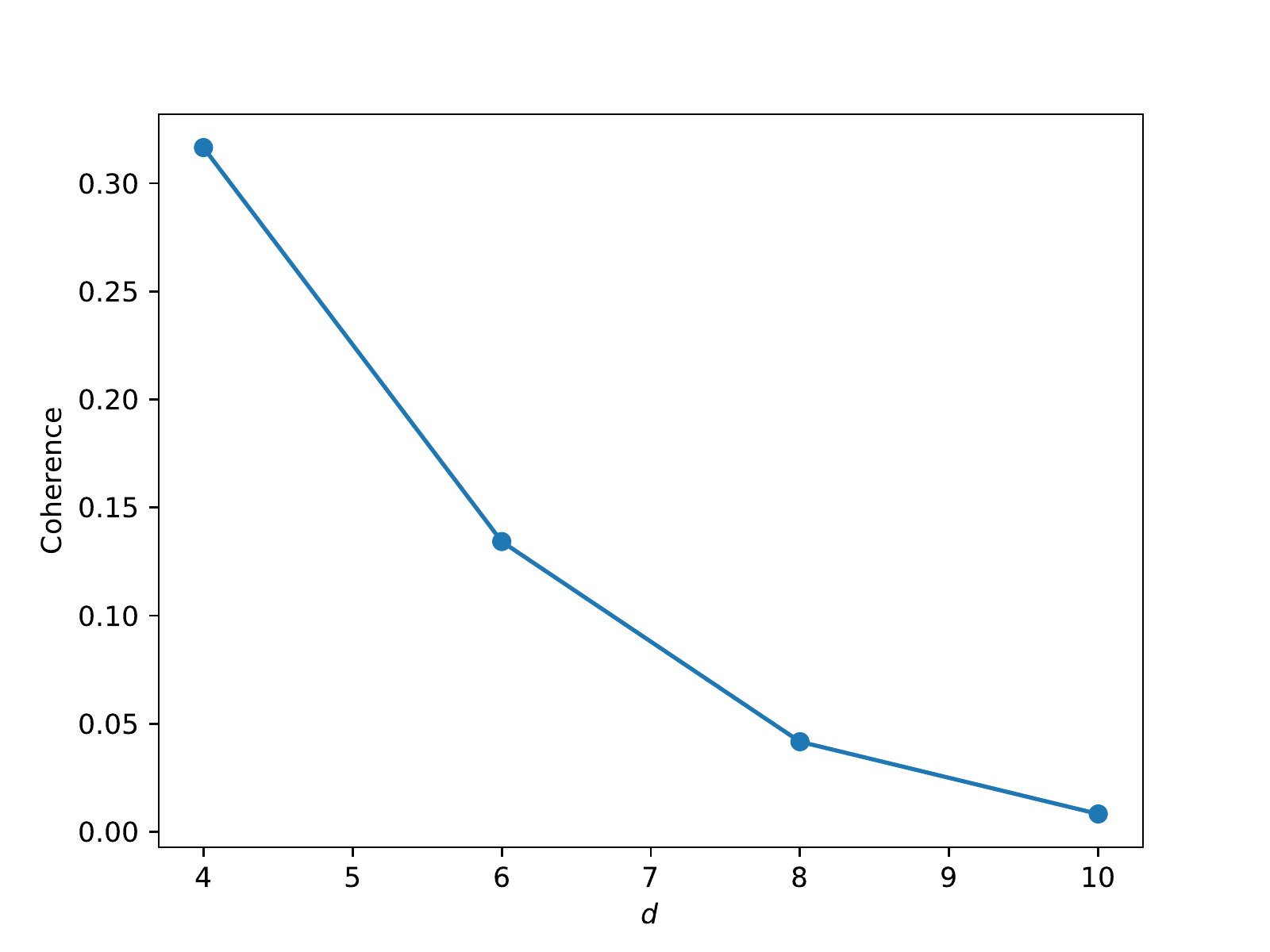}
    \caption{Decrease in coherence, given by the off diagonal component of the density matrix, of a SCS with increase in qudit dimension $d=2I+1$.}
    \label{fig:catlifetime}
\end{figure}

We study the scaling of the coherence of a SCS with respect to the Hilbert space dimension of a qudit~\cite{WHSK20FrontPhys} and consider dephasing purely due to magnetic field fluctuations, of the state $\ket{I,I}+\ket{I,-I}$  for different values of $I$.
The density matrix of such a state consists only of four elements: two diagonal elements corresponding to the population in $\ket{I,\pm I}$ and two off-diagonal components that quantify the coherence of the superposition state.
In Fig.~\ref{fig:catlifetime}, we plot the magnitude of the off-diagonal component after a fixed time of 1 ms under an amplified dephasing at rate of $\Gamma_{\text m} = 1$~kHz ($\Gamma_\text m\propto1/T_2^*$).
We note that coherence decreases with the dimension $d=2I+1$ of the nuclear spin and results in a linear decrease in coherence time as the dimension of a qudit increases.
Such a trade-off between lifetime and the macroscopicity of a SCS is similarly found in other systems such as harmonic oscillators and maximally entangled GHZ states~\cite{HMP+97PRL}.
However, compared to other systems, a nuclear spin, with naturally long coherence times, can help increase the lifetime of practically achievable SCSs.

\section{Converting two-axis counter twisting to one-axis twisting}
\subsection{Two-axis counter twisting}

\begin{figure*}
    \centering
    \includegraphics[width = \linewidth]{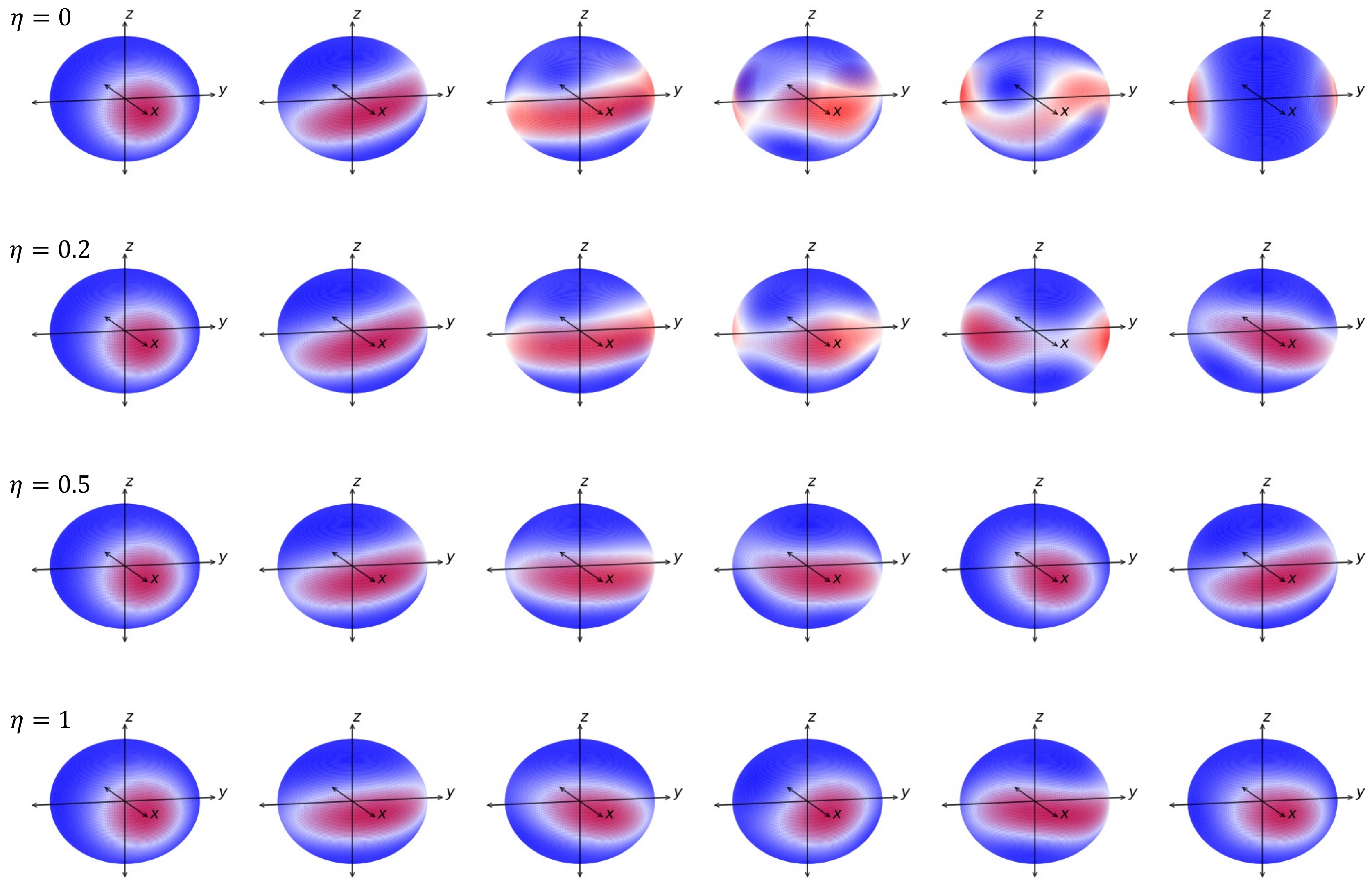}
    \caption{Nuclear spin squeezing under non-linear quadrupole interaction for different values of asymmetry parameter $\eta$. The quadrupole interaction can produce spin cat states (SCSs) for $\eta = 0$, however, for higher values of $\eta$, only spin squeezed states are formed. One can note that the period of non-linear evolution decreases as $\eta$ increases, for e.g.~the coherent state reappears faster for $\eta=1$ (third figure from right), in comparison to $\eta=0.5$ (second figure from right).}
    \label{fig:tact}
\end{figure*}

Substituting $\mathbb{I}^2=\hat I_{x'}^2 + \hat I_{y'}^2 + \hat I_{z'}^2 $ in Eq.~\eqref{eq:SMquadrupole}, the quadrupole interaction can be described as
\begin{equation}
    \hat H_\text{q} = \omega_\text{q} \left(1-\frac{\eta}{3}\right)\left[ \hat I_{z'}^2-a\hat I_{y'}^2 \right]; a =\left(\frac{2\eta}{3-\eta}\right),
\end{equation}
which is equivalent to one-axis twisting (OAT) when $\eta = 0$, two-axis counter twisting (TACT) when $\eta=1$, and partial TACT otherwise. 
Under TACT, the nuclear spin is simultaneously twisted about two axes, which increases the squeezing speed 
and the nuclear spin returns to a coherent state faster than in the case of OAT (as can be noted in Fig.~\ref{fig:tact} for $\eta = 1$). 
During this, the spin does not spread over the equator of the Bloch sphere due to counter twisting, which prevents interference and cat-state formation. 

The quadrupole interaction can only generate squeezed states for $\eta\neq0$, as can be seen in Fig.~\ref{fig:tact}. For small $\eta$, partial TACT can create cat-like states, as shown for $\eta=0.2$ in Fig.~\ref{fig:tact}; however, its two components are not spin coherent states and the separation between them is smaller than that for a macroscopic SCS. 
In practice, there is little to no control over the in-situ EFG generated by the lattice distortions, and $\eta$ depends on several experimental factors. 
Thus, quadrupole interaction alone cannot be used for realizing SCSs.

\subsection{Adding a linear interaction}

\begin{figure*}
    \centering
    \includegraphics[width=\linewidth]{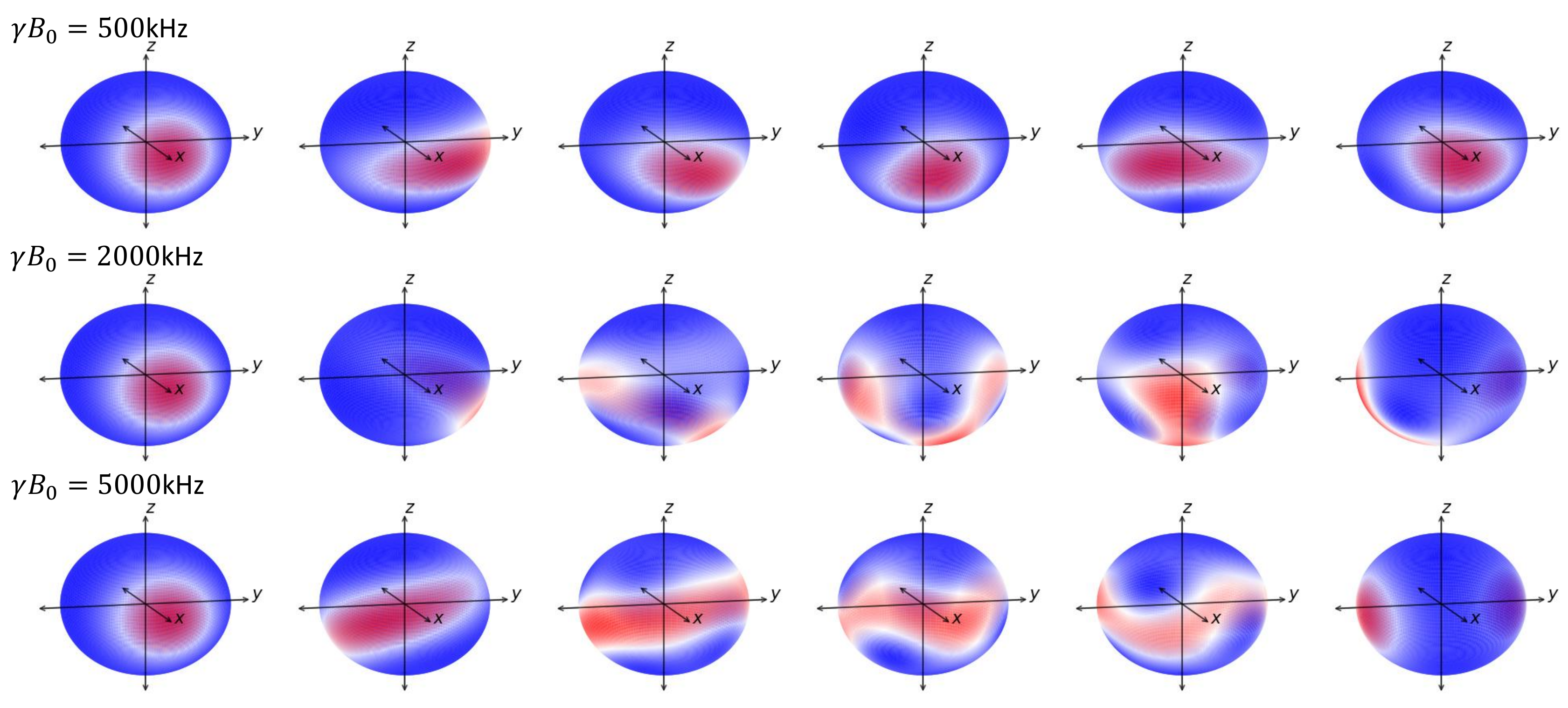}
    \caption{Dynamics when $\eta = 1$ for different Zeeman interaction strength $\gamma B_0$ and quadrupole strength $\omega_\text q = 2\pi\times 40\approx250$~kHz. SCSs are formed by converting two-axis counter twisting to one-axis twisting by adding a large linear interaction induced by the Zeeman interaction ($\gamma B_0\gg\omega_\text q$). }
    \label{fig:oat-supplement}
\end{figure*}

We convert TACT to OAT by adding a large linear term, introduced here by the Zeeman interaction, that ``washes out" TACT. 
To understand this, we note that spin squeezing is caused by non-linear transitions induced by $\hat I_{z'}^2$ and $\hat{I}_{y'}^2$, and,
for the simple case of $\{x',y',z'\}\equiv \{x, y, z\}$, $\hat I_z^2$ causes non-linear phase shifts and $\hat I_y^2$ causes quadrupolar transitions between different $\hat I_z$ energy eigenstates  with $\Delta m_I =\pm 2$.
TACT can be converted to OAT if $\hat{I}_{y}^2$-induced quadrupolar transitions can be prevented.
This is achieved when the Zeeman splitting is much larger than quadrupole coupling, i.e.~$\gamma\bm B_0\gg \omega_\text{q}$, and non-linear transitions induced by the relatively small $\hat I_{y}^2$ term are prevented by the large energy level splitting.
In such a case, only the component of the quadrupole interaction parallel to Zeeman interaction generates squeezing, which results in an effective one-axis twisting, regardless of the value of $\eta$.
Figure \ref{fig:oat-supplement} shows nuclear spin squeezing under different strengths of quadrupole interaction, where we note that SCSs are formed only for a very large value of $\gamma B_0$.

\subsection{Effect of quadrupolar orientation}
Above, we analyzed the special case of $\{x',y',z'\}\equiv \{x, y, z\}$; however, the conversion of TACT to OAT holds for any set of Euler angles $\{\delta,\mu,\nu\}$ and our scheme can create SCSs in-situ for any configuration of the nuclear quadrupolar interaction.
In the presence of a large Zeeman interaction, effective spin squeezing is only caused by $Q_{zz}$~\eqref{eq:SMHamiltonian} and any other non-linear effect is washed-out.
$Q_{zz}$ is always non-zero except when the principal axis $z'$ is perpendicular to the Zeeman interaction, i.e.~$\mu=\pi/2$~\eqref{eq:SMeulerz} \textit{and} the asymmetry parameter $\eta = 0$, resulting in a quadrupole interaction similar to $\hat I_x^2$. 
In this corner case of OAT, an initial nuclear spin coherent state can form a spin SCS without any additional control.

\end{document}